# MIMO Detection Algorithms for High Data Rate Wireless Transmission

Nirmalendu Bikas Sinha, R.Bera and M.Mitra

**Abstract**— Motivated by MIMO broad-band fading channel model, in this section a comparative study is presented regarding various uncoded adaptive and non-adaptive MIMO detection algorithms with respect to BER/PER performance, and hardware complexity. All the simulations are conducted within MIMO-OFDM framework and with a packet structure similar to that of IEEE 802.11a/g standard. As the comparison results show, the RLS algorithm appears to be an affordable solution for wideband MIMO system targeting at Giga-bit wireless transmission. So MIMO can overcome huge processing power required for MIMO detection by using optimizing channel coding and MIMO detection.

——————————— ◆ ———————————

## 1. INTRODUCTION

In recent years, multiple input - multiple output (MIMO) based wireless communications has received widespread attention in the communication community. To date, a majority of the work in this area has been of a theoretical nature [1], [2], [3] and little attention has been paid to the implementation requirements of MIMO systems. Recently the UCLA Wireless Integrated Research (WISR) group embarked on a project to develop a wideband (25MHz) real-time MIMO-OFDM test bed at 5.2GHz RF. The ultimate objective is to develop both system solution and novel VLSI architecture to enable real-time Gigabps indoor wireless communications. One of the challenges in building a wideband MIMO system is the tremendous processing power required at the receiver side. While coded MIMO schemes offer better performance than separate channel coding and modulation scheme by fully exploring the trade-off between multiplexing and diversity [4], its hardware complexity can be practically formidable, especially for wideband system with more than 4 antennas on both transmitter and receiver sides. On the other hand, it's much easier to find a VLSI solution using traditional channel coding schemes such as convolution code and Turbo code for data rate of hundreds of Mbps. For this reason, we start off by considering the uncoded MIMO schemes, also called spatial multiplexing as shown in Fig. 1, and carry out a side-by-side comparative study to evaluate a number of uncoded MIMO detection algorithms from both performance and implementation point of view.

————————————————

- *Prof. Nirmalendu Bikas Sinha, corresponding author is with the Department of ECE and EIE , College of Engineering & Management, Kolaghat, K.T.P.P Township, Purba- Medinipur, 721171, W.B., India.*

- *Dr. R. Bera is with the S.M.I.T, SikkimManipal University, Majitar, Rangpo, East Sikkim, 737132.*

- *Dr. M.Mitra is With the Bengal Engineering and science University, Shibpur, Howrah, India .*

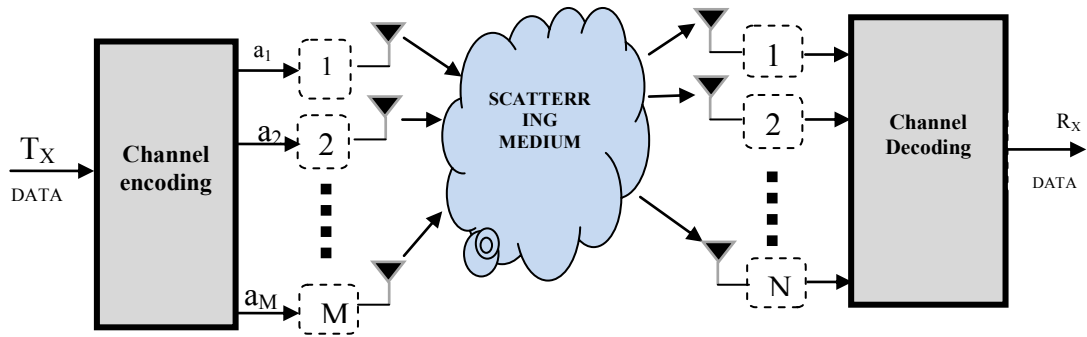

Fig. 1, Block diagram of spatial multiplexing system.

## 2. MIMO DETECTION FOR FLAT-FADING CHANNEL

### A) MIMO Channel Model

MIMO systems are an extension of smart antennas systems. Traditional smart antenna systems employ multiple antennas at the receiver, whereas in a general MIMO system multiple antennas are employed both at the transmitter and the receiver. The addition of multiple antennas at the transmitter combined with advanced signal processing algorithms at the transmitter and the receiver yields significant advantage over traditional smart antenna systems - both in terms of capacity and diversity advantage. A MIMO channel is a wireless link between M transmits and N receive antennas. It consists of MN elements that represent the MIMO channel coefficients. The multiple transmit and receive antennas could belong to a single user modem or it could be distributed among different users. The later configuration is called distributed MIMO and cooperative communications. Statistical MIMO channel models offer flexibility in selecting the channel parameters, temporal and spatial correlations. MIMO channel simulation tools are implemented based on these models. Several statistical MIMO channel models were proposed in [5] and [6]. Both models introduced spatial correlation by multiplying a matrix of uncorrelated random variables by a square root of a covariance matrix and both are based on similar assumptions. However, they differ in their approach. In [7], the authors validate the statistical model of [5] based on measurements in microcells and microcells. They showed that the eigen value distribution of the model matches the measurements. Fig.2 (a), (b), (c) and (d) shows conceptual diagram of existing technology, smart antenna system and MIMO channels respectively.

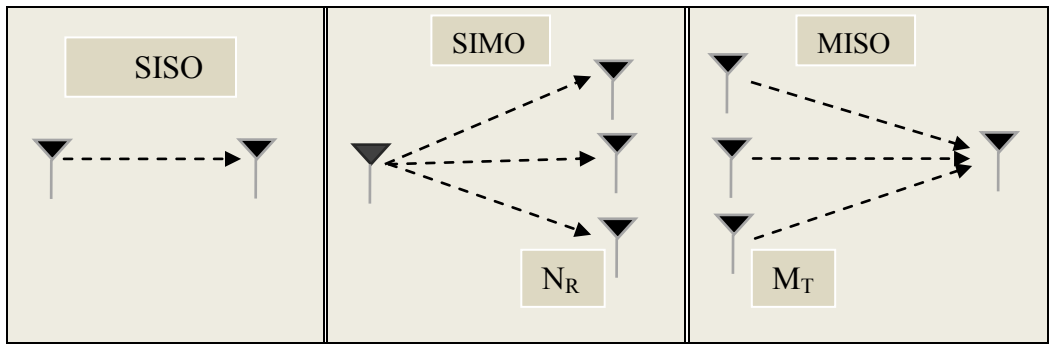

**Fig. 2 (a) Existing technology, (b) & (c) Smart antenna system**

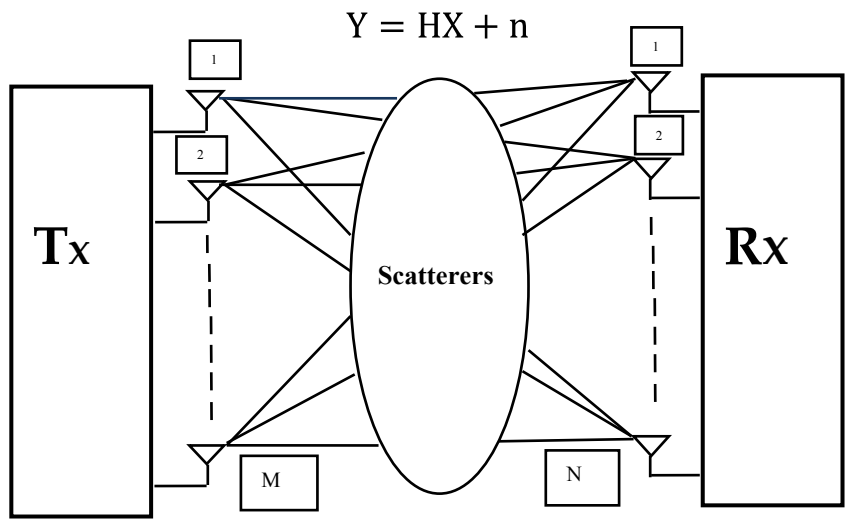

Fig. 1(d) A MIMO wireless channel

### 3. Performance Analysis of different MIMO detectors

A straightforward approach to recover x from y is to use an M X N weight matrix W to linearly combine the elements of y to estimate x, i.e. $\hat{x} = W_y$.

#### 3.1 *Maximum Likelihood (ML):*

The ML receiver performs optimum vector decoding and is optimal in the sense of minimizing the error probability. ML receiver is a method that compares the received signals with all possible transmitted signal vector which is modified by channel matrix H and estimates transmit symbol vector x according to the Maximum Likelihood principle, which is shown as:

$$\hat{x} = \arg_{x_k \in \{x_1, x_2 \ldots x_N\}} min\|r - Hx_k\|^2 \quad \ldots(1)$$

Where the minimization is performed over all possible transmit estimated vector symbols *x*. Although ML detection offers optimal error performance, it suffers from complexity issues. It has exponential complexity in the sense that the receiver has to consider $|A|^M$ possible symbols for an M transmitter antenna system with A is the modulation constellation.

**3.2 V-BLAST Zero Forcing (ZF) characteristic:**

We can reduce the decoding complexity of the ML receiver significantly by employing linear receiver front-ends to separate the transmitted data streams, and then independently decode each of the streams. Simple linear receiver with low computational complexity and suffers from noise enhancement. It works best with high SNR. The solution of the ZF is given by:

$$\hat{x} = (H^*H)^{-1}Hx = H^+x \dots (2)$$

Where, ( )$^+$represents the pseudo-inverse. The ZF receiver converts the joint decoding problem into M single stream decoding problems thereby significantly reducing receiver complexity. This complexity reduction comes, however, at the expense of noise enhancement which in general results in a significant performance degradation (compared to the ML decoder). The diversity order achieved by each of the individual data streams equals N - M + 1.

**3.3. V-BLAST with Minimum Mean Square Error (MMSE):**

The MMSE receiver suppresses both the interference and noise components, whereas the ZF receiver removes only the interference components. This implies that the mean square error between the transmitted symbols and the estimate of the receiver is minimized. Hence, MMSE is superior to ZF in the presence of noise. Some of the important characteristics of MMSE detector are simple linear receiver, superior performance to ZF and at Low SNR, MMSE becomes matched filter. Also at high SNR, MMSE becomes Zero-Forcing. MMSE receiver gives a solution of:

$$\hat{x} = D \cdot x = \left(\frac{1}{SNR}I_{N_R} + H^HH\right)^{-1} \cdot H^Hx \dots (3)$$

At low SNR, MMSE becomes ZF:

$$\left(\frac{1}{SNR}I_{M_R} + H^HH\right)^{-1}H^H \approx \frac{1}{SNR}H^H \dots (4)$$

At high SNR, MMSE becomes ZF:

$$\hat{x} = D \cdot x = \left(\frac{1}{SNR}I_{M_T} + H^HH\right)^{-1}H^H \approx (H^HH)^{-1}H^H \dots (5)$$

i.e., the MMSE receiver approaches the ZF receiver and therefore realizes (N-M + 1)$^{th}$ order diversity for each data stream.

**3.4 .V-BLAST with Maximal Ratio Combining (MRC):**

MRC combines the information from all the received branches in order to maximize the ratio of signal to noise power, which gives it its name. MRC works by weighting each branch with a complex factor and then adding up the branches, MRC is intuitively appealing: the total SNR is achieved by simply adding up the branch SNRs when the appropriate weighting coefficients are used.
BER for MRC in Rayleigh fading channel (1x2) with BPSK modulation,

$$P_e\ MRC = P_{MRC}^2[1 + 2(1 - p_{MRC})] \dots (6)$$

$$p_{MRC} = \frac{1}{2} - \frac{1}{2}\left(1 + \frac{1}{E_b/N_o}\right)^{-1/2} \dots (7)$$

**3.5. STBC (space-time block codes):**

STBC is a class of linear coding for MIMO systems that aims to maximize the system diversity gain rather than the data rate. A very popular STBC for a two transmit antennas setup was developed by Alamouti, which is illustrated in Fig.3. It is designed for 2x2 MIMO systems and its simplicity and high frequency have led to its wide adoption in MIMO systems. In this scheme orthogonal signals are transmitted from each antenna, which greatly simplifies receiver design.

This particular scheme is restricted to using M = 2 antennas at the transmitter but can any number of receive antennas N .Two QAM symbols $S_1$ and $S_2$ for transmission by the Alamouti scheme are encoded in both the space and time domain at the two transmitter antennas over the consecutive symbol periods as shown in equation( 20). The information bits are first modulated using a modulation scheme (for example QPSK). The encoder then takes a block of two modulated symbols $s_1$ and $s_2$ in each encoding operation and gives to the transmit antennas according to the code matrix,

$$S = [\mathbf{s_1} \quad \mathbf{s_2}] = \begin{bmatrix} s_1 & -s_2^* \\ s_2 & s_1^* \end{bmatrix} \ldots \ldots (8)$$

The code matrix has the following property

$$S.S^H = \begin{bmatrix} |x_1|^2 + |x_0|^2 & 0 \\ 0 & |x_1|^2 + |x_0|^2 \end{bmatrix}$$
$$= (|x_1|^2 + |x_0|^2)I_2 \ldots (9)$$

Where $I_2$ is the 2x2 identity matrix.

In the above matrix the first column represents the first transmission periods and the second column, the second transmission period. The first row corresponds to the symbols transmitted from the first antenna and second row corresponds to the symbols transmitted from the second antenna. It means that during the symbol period, the first antenna transmits $s_1$ and second antenna $s_2$. During the second symbol period, the first antenna transmits $-s_2^*$ and the second antenna transmits $s_1^*$ being the complex conjugate of $s_1$. This implies that we are transmitting both in space (across two antennas) and time (two transmission intervals). This is space time coding. Hence, $\mathbf{S_1}= [s_1 \ -s_2^*]$ and $\mathbf{S_2}= [s_2 \ s_1^*]$ Moreover a close look reveals that sequences are orthogonal over a frame interval, since the inner product of the sequences $\mathbf{S_1}$ and $\mathbf{S_2}$ is zero, i.e.

$$\mathbf{S_1}.\mathbf{S_2} = s_1 s_2^* - s_2^* s_1 = 0 \ldots \ldots (10)$$

$$P_e \, STBC = P_{STBC}^2 [1 + 2(1 - p_{STBC})] \ldots (11)$$

$$p_{STBC} = \frac{1}{2} - \frac{1}{2}\left(1 + \frac{2}{E_b/N_o}\right)^{-1/2} \ldots \ldots (12)$$

In a fast fading channel, the BER is of primary interest since the channel varies every symbol time; while in a slow fading situation, the FER (Frequency error rate) is more important because channel stays the same for a frame.

*3.6. Linear Adaptive MIMO Detection*

Instead of assuming known channel matrix **H**, which usually requires channel probing before each transmission and then calculating **W** in a bursty manner, adaptive algorithms estimate **W** directly through iteration via the use of a known training sequence at the beginning of each transmission.

*A) Least Mean-Square (LMS):* LMS is an estimate of the steepest descent algorithm [5] and updates **W** according to

$$W_i = W_{i-1} + \mu[x_i - W_{i-1} y_i]y_i^* \ldots \ldots (13)$$

Where μ is the update step size. For LMS to to converge in the mean-squared sense, i.e, $j_i \rightarrow j_\infty$, μ needs to satisfy $0 < \mu < 2/\lambda_{max}$. Where $\lambda_{max}$ is the largest eigen value of $R_y = \sigma_x^2 HH^* + \sigma_v^2 I$. Therefore, the convergence of LMS depends on both channel condition and signal to noise ratio at the input of the receiver. The final residual error also depends on the value of μ.

*B) Recursive Least-Squares (RLS):* RLS is the recursive solution to the exponentially weighted least-squares (LS) problem [5]. The recursive optimal solution at time instant *i* is

$$W_i = W_{i-1} + [x_i - W_{i-1} y_i]y_i^* P_i \ldots \ldots (14)$$

$$P_i = \lambda^{-1}\left[P_{i-1} - \frac{\lambda^{-1} P_{i-1} y_i y_i^* P_{i-1}}{1 + \lambda^{-1} y_i y_i^* P_{i-1}}\right] \ldots (15)$$

$0 \ll \lambda < 1$ is the exponential forgetting factor, and $P_i = (\sum_{K=0}^{i} \lambda^{i-k} y_i y_i^*)^{-1}$ is the inverse of the weighted correlation matrix of $\mathbf{y}_i$ with initial condition $\boldsymbol{P_{-1} = \pi_0 I}$

The scalar $\boldsymbol{\pi_0}$ is usually a large positive number and $\lambda$ is very close to 1. Compared to the stochastic estimation problems given previously which require the signal statistics such as correlation matrix, LS problem is deterministic [8], [9]. Therefore, RLS can be used to find the LS solution to a non-stationary process, or simply said, RLS can track nonstationary process in the LS sense. When $\mathbf{x}_i$, **H**, and $\mathbf{v}_i$ are all stationary, it is the weighted time-average estimate to MMSE as $i \rightarrow \infty$ if **Rxy** and **Ry** are replaced by $\sum_{K=0}^{i} \lambda^{i-k} x_k y_k^*$ and $\sum_{K=0}^{i} \lambda^{i-k} y_k y_k^*$ respectively.

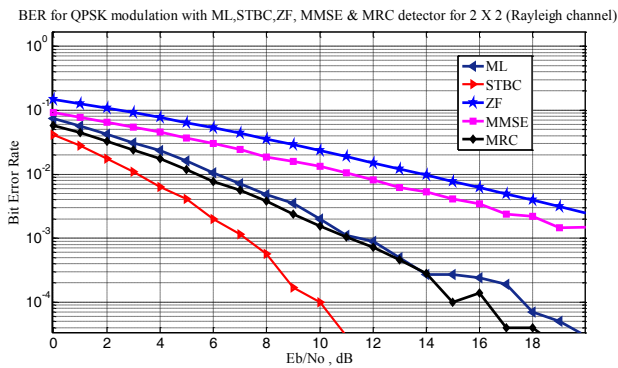

Fig.3, Performance curves for different linear detectors (ML, STBC, ZF, MMSE, and MRC) in 2×2 MIMO - V-BLAST system in a slow fading channel.

Fig. 3 shows all the simulation results. At a certain Bit Error Rate point, BER=0.001, there is approximately 2.3dB SNR difference between the V-BLAST SIC system with ZF detectors and the V-BLAST SIC ordering system. The difference is smaller than what we expected. The performance curves of these two systems are close to each other, especially when the SNR is low, but the gap gets larger when the SNR gets higher. When the SNR is low, which means the noise is large, the post detection SNR is mainly affected by the noise, thus we will not see a big difference between the SIC system with or without ordering. When the SNR gets higher, the post detection SNR is mainly affected by the channel matrix H. The post detection SNR of a stream will be of great difference when the stream is suffering from a deep fading, it is sensitive to the channel characteristic. If there are more antennas in the transmitter, which means there are more stages, or the channel condition is more complicated, we will observe more improvement from using the ordering strategy. If we use a MRC detector instead of a ZF detector at the first stage of the V-BLAST SIC ordering system, we will have a gain of 12.3dB; this gain comes from the joint ML detector. When the BER is equal to 0.001, we need SNR=3dB in the general V-BLAST system with the ML detector, and we need SNR=4.6dB in the SIC ordering system with the ML detector at the first stage. That is a difference of only 1.6dB, thus we can use the SIC ordering system with first stage ML instead of the general V-BLAST scheme since these two schemes perform similarly, and we do not need to code across the transmitting antennas.

*C) PER Performance*

The PER performance curves are shown in Fig. 4. ZF and MMSE yield very close PER performance, and similarly, ZFVBLAST and MMSE-VBLAST (except for M = N > 1). For this reason, the curves in Fig. 4 only illustrate MMSE and MMSE-VBLAST. The performance of VBLAST is consistently better than ZF/MMSE since for PER to reach below 100%, the SNR is already sufficiently high for infrequent error propagation. At α= 1, the PER for MMSE increases with the number of antennas as compared to the roughly overlapped curves previously observed in the BER plot. For $\tau_{rms}$ = 50ns, the channel selectivity leads to a degradation in PER compared to flat-fading channel. This is because for each packet, it's more likely to see bit errors caused by a deep null in the channel frequency response (corresponding to lower SNR), than in channels with smaller $\tau_{rms}$. This is readily mitigated via interleaved channel coding techniques. On the other hand, the BER stays the same for different $\tau_{rms}$ as long as the cyclic prefix is sufficiently long compared to $\tau_{rms}$. In our simulations, we have used a cyclic prefix length of 16 × 40ns= 640ns, long enough for $\tau_{rms}$ = 50ns.

*D) Convergence of LMS/RLS*

LMS and RLS are resursive alternatives to the matrix inversion-based solutions. Under the simulation environments, the BER/PER performance of RLS will ultimately converge to the MMSE results shown before with sufficient training while that of LMS should get very close when *µ* is very small. What's key here is the convergence speed, as shown in Fig. 5 in the form of ensemble average BER learning curves. As expected, at the same SNR, LMS converges much slower than RLS except for 1 × 1 case where the learning curves actually overlap with each other. The required training length depends on various factors including number of antennas, SNR, and updating factor *µ* or *λ* For RLS, the training length is roughly on the order of *M · N*. For LMS, it's about 10 times longer. At higher SNR, it takes RLS longer to converge because the learning curve at higher SNR has a deeper BER floor to reach while the initial (also fastest) learning slope for different SNR is similar. Larger *µ* or smaller *λ* can increase the speed of convergence for LMS and RLS, respectively. But on the other hand, they tend to adversely affect the final convergence performance at the same time. The convergence of LMS also depends on the channel condition number (eigenvalue spread). RLS guarantees the convergence at the price of higher hardware complexity and less robustness to quantization effects [8], [9]. Similar to the BER performance, the effect of channel delay spread on the convergence of LMS/RLS is negligible when the cyclic prefix is long enough.

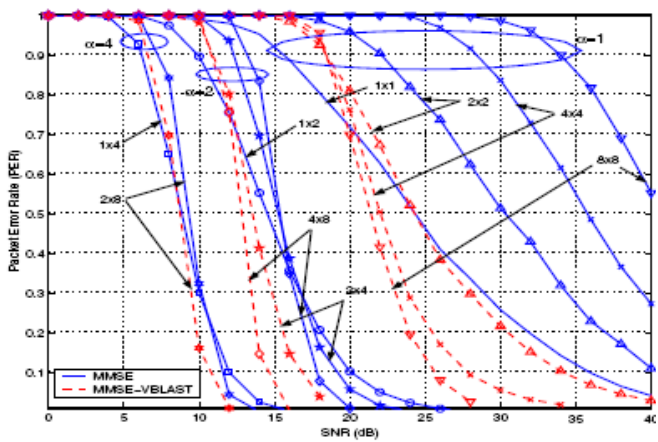

Fig. 4. PER performance of MMSE and MMSE-VBLAST

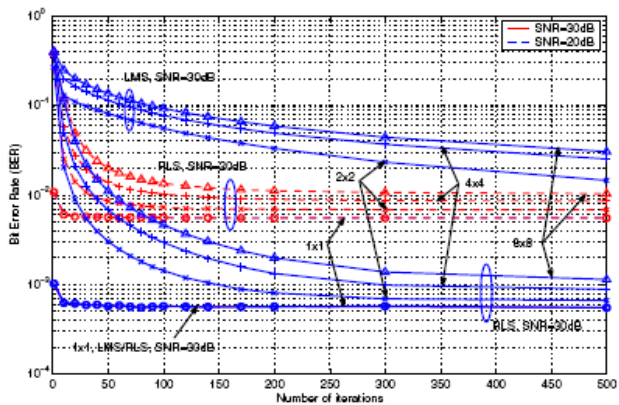

Fig. 5. The learning curves of LMS ($\mu$ = 0.02) and RLS ($\lambda$ = 0.99)

## CONCLUSION

In this paper, we have provided an overview of various MIMO detection algorithms for spatial multiplexing systems. The simulated performance of these algorithms is compared, and this comparison is further extended to a first order estimation of their hardware costs. From these comparisons, it is observed that VBLAST generally outperforms ZF/MMSE, at the cost of significantly higher implementation complexity. In fact, ZF-VBLAST is only slightly better than MMSE (2 ~ 3dB at uncoded BER of 10−3). The advantages of VBLAST over ZF/MMSE become much less significant when $\alpha > 1$ because of the antenna diversity gain. MMSE-VBLAST performs much better than ZF-VBLAST. However, in practice, inaccurate estimates as well as the channel matrix itself tend to reduce this gain. The study shows that RLS has a much lower computation intensity than ZF/MMSE/VBLAST and achieves the performance of MMSE with sufficient training. This is done by spreading out the computation through multiple iterations. RLS is superior to LMS in terms of convergence speed, and its hardware cost is on the same order as LMS. Compared to ZF/MMSE/VBLAST, RLS doesn't require explicit channel information and subsequent matrix inversion, and can be implemented using the QR-decomposition based systolic array architecture [8] [9]. Therefore, it can be concluded that RLS presents the best performance to complexity metric among the surveyed algorithms for Giga-bps MIMO wireless systems. Based on these findings, RLS has been chosen for the MIMO detection in the UCLA MIMO-OFDM testbed [10], [11].

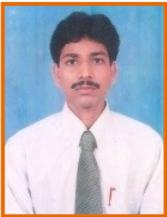

**Prof. Nirmalendu Bikas Sinha** received the B.Sc (Honours in Physics), B. Tech, M. Tech degrees in Radio-Physics and Electronics from Calcutta University, Calcutta,India,in1996,1999 and 2001, respectively. He is currently working towards the Ph.D degree in Electronics and Telecommunication Engineering at BESU. Since 2003, he has been associated with the College of Engineering and Management, Kolaghat. W.B, India where he is currently an Asst.Professor is with the department of Electronics & Communication Engineering & Electronics & Instrumentation Engineering. His current research Interests are in the area of signal processing for high-speed digital communications, signal detection, MIMO, multiuser communications,Microwave /Millimeter wave based Broadband Wireless Mobile Communication ,semiconductor Devices, Remote Sensing, Digital Radar, RCS Imaging, and Wireless 4G communication.  He has published large number of papers in different international Conference, proceedings and journals.He is presently the editor and reviewers in different international journals. E-mail: nbsinha@yahoo.co.in

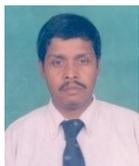

**Dr. Rabindranath Bera** is a professor and Dean (R&D), HOD in Sikkim Manipal University and Ex-reader of Calcutta University, India. B.Tech, M.Tech and  Ph.D.degrees from Institute of Radio-Physics and Electronics, Calcutta University. His research areas are in the field of Digital Radar, RCS Imaging, Wireless 4G Communications, Radiometric remote sensing. He has published large number of papers in different national and international Conference and journals.

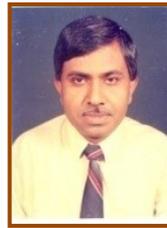

**Dr. Monojit Mitra** is an Assistant Professor in the Department of Electronics & Telecommunication Engineering of Bengal Engineering & Science University, Shibpur. He obtained his B.Tech, M.Tech & Ph. D .degrees from Calcutta University. His research areas are in the field of Microwave & Microelectronics, especially in the fabrication of high frequency solid state devices like IMPATT. He has published large number of papers in different national and international journals. He has handled sponsored research projects of DOE and DRDO. He is a member of IETE (I) and Institution of Engineers (I) society.